\documentclass[rnote]{aa}
\usepackage{amsmath,txfonts,epsf,epsfig,subfig,natbib} 

\bibpunct{(}{)}{;}{a}{}{,} 

\begin{document}


\title{Effect of the toroidal magnetic field on the runaway instability of relativistic tori}

\author{Jaroslav Hamersk\'y\inst{1,2}\thanks{\email{hamersky@astro.cas.cz}} \and Vladim\'{\i}r Karas\inst{1}}
\institute{Astronomical Institute, Academy of Sciences, Bo\v{c}n\'{\i} II 1401, CZ-14100 Prague, Czech~Republic \and
Charles University in Prague, Faculty of Mathematics and Physics, V Hole\v{s}ovi\v{c}k\'ach 2, CZ-18000 Prague, Czech Republic}
\date{Received 18 April 2013; Accepted 16 May 2013}

\abstract{}{Runaway instability operates in fluid tori around black holes.
It affects systems close to the critical (cusp overflowing) configuration.
The runaway effect depends on the radial profile $l(R)$ of the angular momentum 
distribution of the fluid, on the dimension-less spin $a$ of the central black hole
($|a|\leq1$), and other factors, such as self-gravity. Previously it was 
demonstrated that, for the power-law dependence of the
radial angular momentum profile, $l(R)\propto R^q$, 
non-magnetized tori always become runaway stable for a sufficiently high 
positive value of $q$. Here we discuss the role of runaway instability within 
a framework of an axially symmetric model of perfect fluid endowed with a 
purely toroidal magnetic field.}{The gradual accretion
of material over the cusp transfers the mass and angular momentum
onto the black hole, thereby changing the intrinsic parameters of the Kerr metric. We 
studied the effect of the plasma parameter $\beta$ (ratio of gas to magnetic pressure)
and other parameters of the model on the evolution of critical configurations that are just on 
the verge of cusp overflow.}{By contributing to the total pressure, the magnetic field 
causes small departures from the corresponding non-magnetic configuration 
in the early phases of accretion. However, we show that the toroidal magnetic 
component inside an accretion torus does not change the frequency of its oscillations significantly. 
We identify these oscillations as the radial epicyclic mode in our example.
Nevertheless, these weak effects can trigger the runaway instability even
in situations when the purely hydrodynamical regime of the torus is stable. 
On the other hand, in most cases the stable configuration remains unaffected, 
and the initial deviations gradually decay after several orbital periods.
We show examples of the torus evolution depending on the initial magnetization $\beta$,
the slope $q$, and the spin $a$.}{The 
toroidal magnetic field plays a more important role in the early phases of the accretion process until
the perturbed configuration finds a new equilibrium or disappears because of the runaway instability.}
\authorrunning{J. Hamersk\'y \& V. Karas}                                                   
\titlerunning{Runaway instability of magnetized accretion tori}                
\maketitle


\section{Introduction} 
Toroidal equilibria of perfect fluid in permanent rotation were introduced a long time ago
as an initial step 
on the way towards an astrophysically realistic description of accretion of gaseous material
onto a black hole in active galactic nuclei and black hole binaries \citep{fis76,abr78,pug12}.
These axially symmetric and stationary solutions are subject to various types
of instability \citep[e.g.,][]{frag2013}. Here we concentrate on a global type of instability 
caused by an overflow of material over the cusp of a critical equipotential surface 
\citep{dai97,abr98,kor12}. It was suggested that this may lead to specific features that 
should be observable in the radiation emitted from an accreting black-hole system \citep{zan03}. 

The effect of the mentioned instability can be catastrophic under certain conditions. 
In particular, a black-hole torus becomes runaway unstable if the angular momentum 
profile within the torus does not rise sufficiently fast with radius \citep{abr98,lu00}. The 
role of general relativity effects on the runaway mechanism was studied in \citet{fon02}
in the context of gamma-ray burst sources. These authors found that by allowing the 
mass of the black hole to grow by accretion, the disc becomes unstable. However, the 
parameter space of the problem is much richer than what could be taken into account 
in early works. For example, the self-gravity of the fluid tends to act against the
stability of non-accreting tori \citep{goo88,mas98,mon10,kor11}. 
Furthermore, the spin parameter can play a role for accretion onto a rotating black hole.
In astrophysically realistic models, an interplay of mutually competing effects 
have to be taken into account.

The role of magnetic fields is known to be essential for accretion. Even the 
Rayleigh-stable tori \citep{seg75} with a radially increasing profile, 
$\mbox{d}l/\mbox{d}R>0$, become dynamically unstable because of
turbulence in the presence of a weak magnetic field \citep{bal91}. Here we aim to
clarify the simpler question of the global stability of a rotationally symmetric
black-hole accretion tori, taking into account the effect of
a large-scale (organized) magnetic field that obeys the same axial symmetry.
\citet{kom06} has developed a suitable analytical (toy) model of such a magnetized torus
described by a polytropic equation of state in Kerr metric. In this model the
magnetic field only enters the equilibrium solution for the torus as an additional 
pressure-like term \citep{pug13}. 

We employed this solution as an initial configuration, 
which we then perturbed and evolved numerically by using a two-dimensional 
numerical scheme \citep[HARM; see][]{gam03}. A complementary approach
in the context of gamma-ray bursts
has been developed in \citet{bar11}, who adopted the same initial configuration
of an axially symmetric magnetized torus, which they evolved taking self-gravity
and neutrino cooling mechanisms into account. Although the basic idea
behind the runaway instability has been well-known since the early papers
\citep[see][]{abr83,wil84} --- it is connected with the existence of the innermost 
stable circular orbit around black holes in general relativity --- an
interplay of different effects makes the discussion rather complex, and so simplified
models have their value for understanding the runaway mechanism in astrophysically 
realistic systems.

Assuming axial symmetry is a useful simplification to explore the origin of 
runaway instability, although it is a far too strong constraint for any  
realistic model of an accreting system. Moreover, a purely toroidal structure of the magnetic field 
and complete negligence of radiative cooling are an over-simplification, which we 
adopt in this paper. However, these assumptions allow us to concentrate on the particular
type of the above-mentioned relativistic instability while non-axisymmetric modes
are suppressed. It also helps us to proceed 
systematically through the parameter space of the model to reveal the
dependence on black-hole spin and the magnetic field strength as they act concurrently
within the relativistic scheme. In this respect our work is complementary to three-dimensional
simulations \citep[e.g.,][]{kor12,mck12}, which are more complete and, 
at the same time, more difficult to comprehend.

In sec.\ \ref{fluidtori} we summarize our approach to magnetized tori 
and the numerical scheme used in our simulations. Then we present
our results by comparing properties of magnetized and non-magnetized accretion tori
that are subject to a weak perturbation from the equilibrium state. 
In sec.\ \ref{discussion} we discuss our results and give a brief conclusion.

\begin{figure}
\resizebox{\hsize}{!}{\subfloat{\includegraphics[width=17cm]{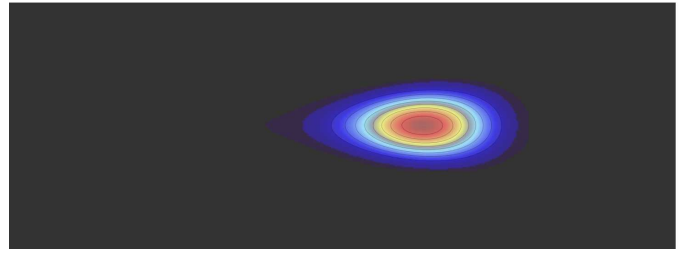}}} \ \ 
\resizebox{\hsize}{!}{ \subfloat{\includegraphics[width=17cm]{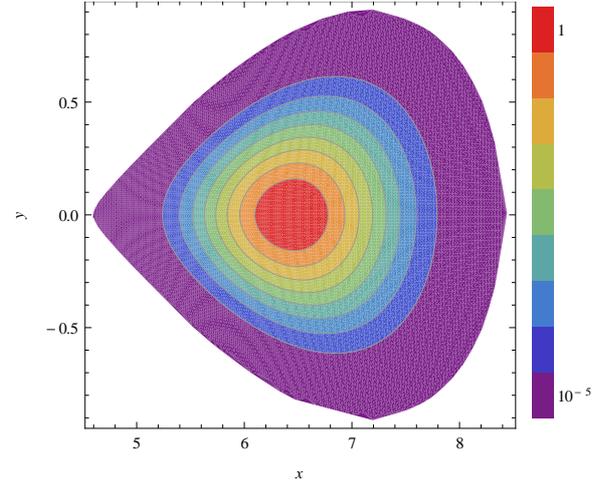}}}
\caption{Stationary distribution of mass across the meridional section of the equilibrium non-magnetised
torus in Kerr metric ($a=0.3$;  the case of a purely hydrodynamical torus).
The top panel shows the shape and the density structure in the poloidal coordinates as 
defined in the HARM code \citep{gam03}, and the same configuration is drawn for Boyer-Lindquist 
coordinates in the bottom panel. See the text for details.}
  \label{Fig1}
\end{figure}

\begin{figure}
\resizebox{\hsize}{!}{\subfloat{\includegraphics[width=8.5cm]{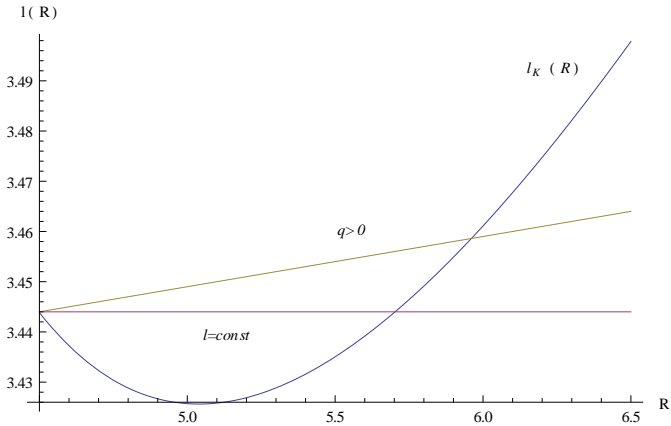}}} 
\caption{Schematic graph of the angular momentum radial profile of the material
inside the accretion torus. At the inner edge ($R_{\rm in}=4.5GM/c^2$) of the matter configuration the angular
momentum is equal to the Keplerian value, $l=l_{\rm K}(R_{\rm in})$. The relativistic Keplerian 
angular momentum distribution first decreases with the radius, passes
through a minimum, and then grows $\propto R^{1/2}$ asymptotically at large radii. The constant value, $l={\rm const}$,
corresponds to the marginally stable configuration, which has been often discussed in the context of
geometrically thick discs. The radially growing profile ($q>0$) improves the stability of the accretion flow.}
  \label{Fig2}
\end{figure}

\begin{figure}
 \resizebox{\hsize}{!}{\includegraphics[width=17cm]{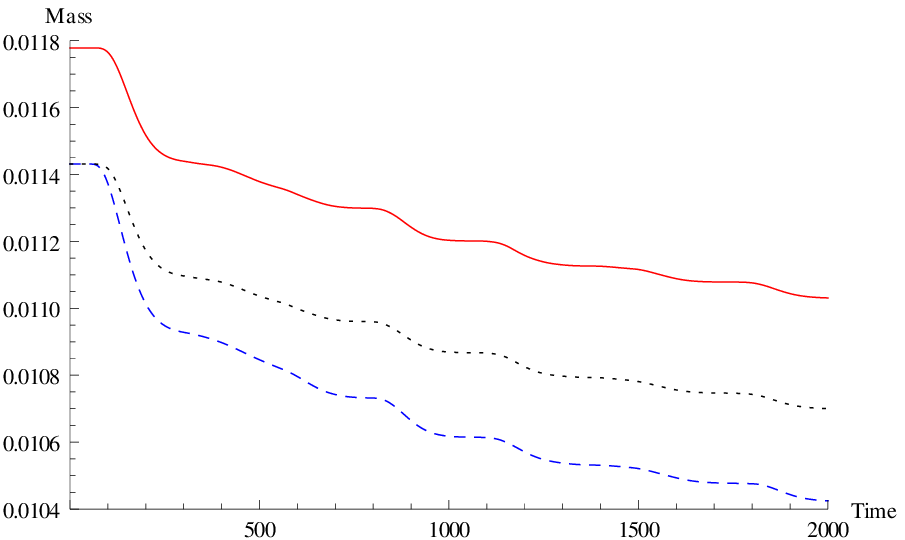}}
 \resizebox{\hsize}{!}{\includegraphics[width=17cm]{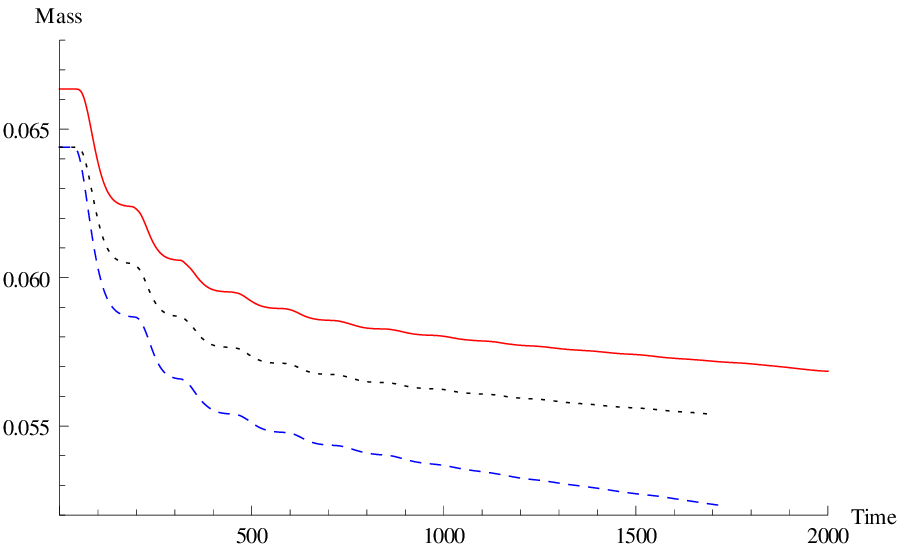}}
 \caption{Torus mass, $M_{\rm d}(t)$, relative to the black-hole mass as a function of time. 
 The initial rapid accretion rate results
 in a drop of $M_{\rm d}$ that becomes partially stabilised during the subsequent evolution. 
 Time is given in dimensionless units of $GM/c^3$. The orbital period is close to its Keplerian
 value near the inner edge, i.e. about $\Delta t(R)\simeq 100$ for the material near $R=R_{\rm in}$.
 Top panel: The case of spin $a=0.1$ is shown for different values of magnetisation parameters
 $\beta = 3$ (dashed), $\beta = 80$ (dotted), and $\beta\rightarrow\infty$
 (i.e. a non-magnetized case; solid line). Bottom panel: as above, but for $a=0.9$.}
 \label{Fig3}
\end{figure}

\section{Axisymmetric accretion of magnetized fluid tori}
\label{fluidtori}
\subsection{Initial configuration}
The magnetized ideal fluid can be described by the energy-momentum tensor 
\citep[e.g.,][]{anile89}
\begin{equation}
T^{\mu\nu}=\left(w+b^2\right)u^{\mu}\,u^{\nu}+P_{\rm g}g^{\mu\nu}-b^{\mu}\,b^{\nu},
\end{equation}
where $w$ is the specific enthalpy, $P_{\rm g}$ is the gas pressure, 
and $b^{\mu}$ is the projection of the magnetic field vector ($b^2=b^{\mu}b_{\mu}$).
 
From the energy-momentum tensor conservation, $T^{\mu\nu}_{\ ;\nu}=0$, it follows for a
purely axially rotating fluid 
(\citeauthor{abr78} \citeyear{abr78,frag2013})
\begin{equation}
\ln |u_t| - \ln |u_{t_{\rm in}}| +\int_0^{P_{\rm g}} \frac{dP}{w}-\int_0^l\frac{\Omega\, dl}{1-\Omega l}+\int_0^{\tilde{P}_{\rm m}}\frac{d\tilde{P}}{\tilde{w}}= 0, \label{AP3}
\end{equation}
where $u_t$ is the covariant component of the four-velocity (subscript ``in'' 
corresponds to the inner edge of the torus), $\Omega=u^{\varphi}/u^t$ 
is the angular velocity and $l=-u_{\varphi}/u_t$ is the angular momentum density.

The specific enthalpy, $w\equiv \rho+P_{\rm g}+U$, can be expressed in terms
of the internal energy density $U$, the rest-mass density $\rho$, and two contributions to the total pressure,
 $P(\rho)=P_{\rm g}+P_{\rm m}$, where $P_{\rm g}$ is the gas (thermodynamical) part 
 and $P_{\rm m}=\frac{1}{2}b^2$ is the magnetic part. The last term introduces 
the effect of the magnetic field in the context of our model and the tilde denotes
$\tilde{w}\equiv {\cal L}w$, $\tilde{P}_{\rm m}\equiv {\cal L}P_{\rm m}$, where 
${\cal L}(r,\vartheta;a)\equiv g_{t\varphi}^2-g_{tt}\,g_{\varphi\varphi}$ is a combination of
metric terms, in our case the known function of radius, latitude, and spin of the Kerr metric. 

By assuming a suitable 
equation of state and the rotation law of the fluid, eq.~\eqref{AP3} can be integrated 
to obtain the structure of equipotential surfaces of the equilibrium configuration.
The last term on the left-hand side of eq.~\eqref{AP3} was assumed to be zero in the 
original solution \citep{abr78}, but \citet{kom06} noticed that its special form 
allows us to take a purely toroidal magnetic field readily into consideration. 
The two contributions are assumed to be proportional to each other. 

Non-vanishing components of the magnetic vector $b^{\mu}$ are
\begin{equation}
b^{\varphi}=\pm \sqrt{\frac{2P_{\rm m}}{\mathcal{A}}},\qquad
b^t=lb^{\varphi}, \label{kap23}
\end{equation}
with $\mathcal{A}(r,\vartheta) \equiv g_{\varphi\varphi}+2lg_{t\varphi}+l^2g_{tt}$.
To solve eq.~(\ref{AP3}) and construct the torus, one needs to constrain the 
angular momentum profile $l(R)$. In this way, we obtain the stationary non-accreting 
configuration, where all viscous effects are neglected. Therefore, accretion can
proceed only due to the relativistic cusp overflow. 

For constant angular momentum density, eq.~\eqref{AP3} can be solved 
analytically as well as in the magnetic case by imposing proportionality 
\begin{equation}
\int_0^{P_{\rm g}} \frac{dP}{w} = C \, \int_0^{\tilde{P}_m}\frac{d\tilde{P}}{\tilde{w}}, \label{IST}
\end{equation}
where $C$ is a constant that sets the mutual relation between hydrodynamic and magnetic 
effects. Since $C$ determines the mutual relation between the hydrodynamic and 
the magnetic pressure components, its value indirectly influences also the 
magnetization ratio $\beta$
(the equipartion state is reached for $\beta$ near unity, whereas $\beta\gg1$ represents a
sub-equipartion magnetic field).
Indeed, the thermodynamic and the magnetic pressure terms and the total pressure 
are directly proportional to each other. This leads to the solution for total pressure in the form
\begin{equation}
P= A {\kappa}^{-3}\left[ \left(  \frac{{u_{t_{\rm in}}}}{{u_t}} \, \exp{\int_0^l\frac{\Omega \mathrm{d}l}{1-\Omega l}} \right)^{\tilde{C}}-1 \right]^4 , \label{Sol5}
\end{equation}
where $A= 0.0039$ and $\tilde{C}=C/(1+C)$ are constants, and $P=\kappa {\rho}^{\gamma}$ 
is the assumed form of equation of state with the polytropic index $\gamma = 4/3$ \citep{frag2013}. 

We focus on critical tori that develop the relativistic cusp at the inner edge, 
$R=R_{\rm in}$, where the angular momentum is equal to the Keplerian angular 
momentum at the corresponding radius in the common equatorial plane of the black 
hole and the torus ($\vartheta=\pi/2$). The mass accretion can either bring
the system out of the critical configuration and stabilise it with an updated set of 
model parameters and a modified profile of angular momentum,
or the accretion process continues in a runaway mode and leads to a complete destruction
of the torus.

\begin{figure*}
\begin{center}
~\hfill~
\includegraphics[angle=0,width=0.4\textwidth]{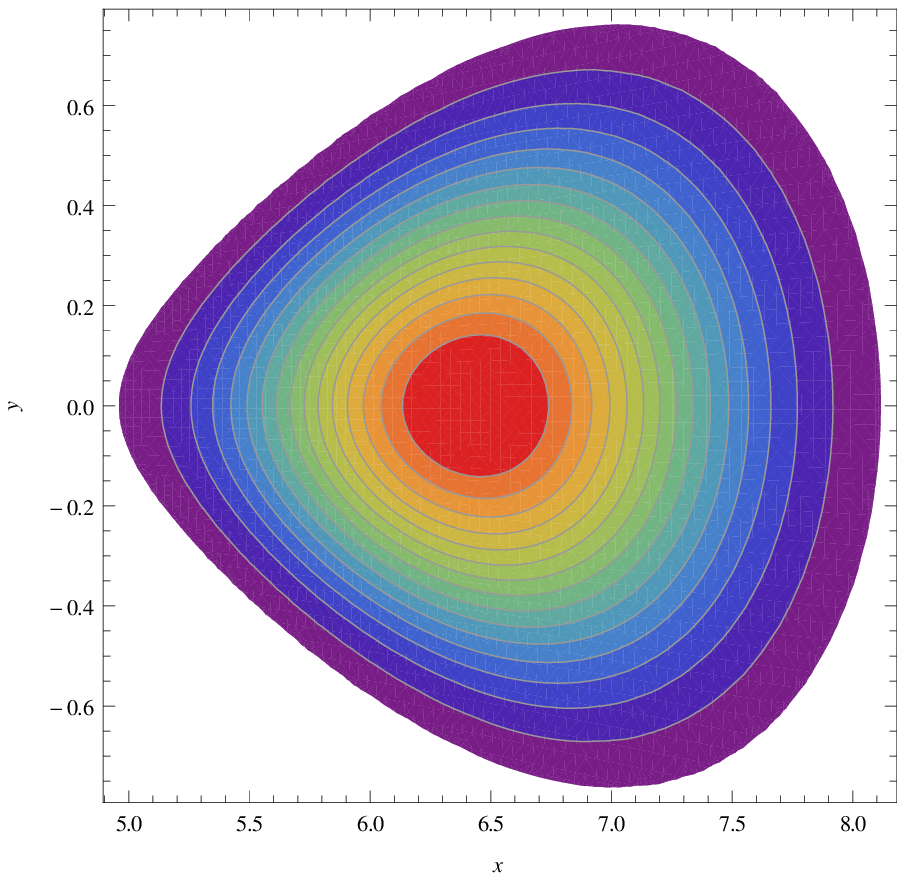}
\hfill
\includegraphics[angle=0,width=0.4\textwidth]{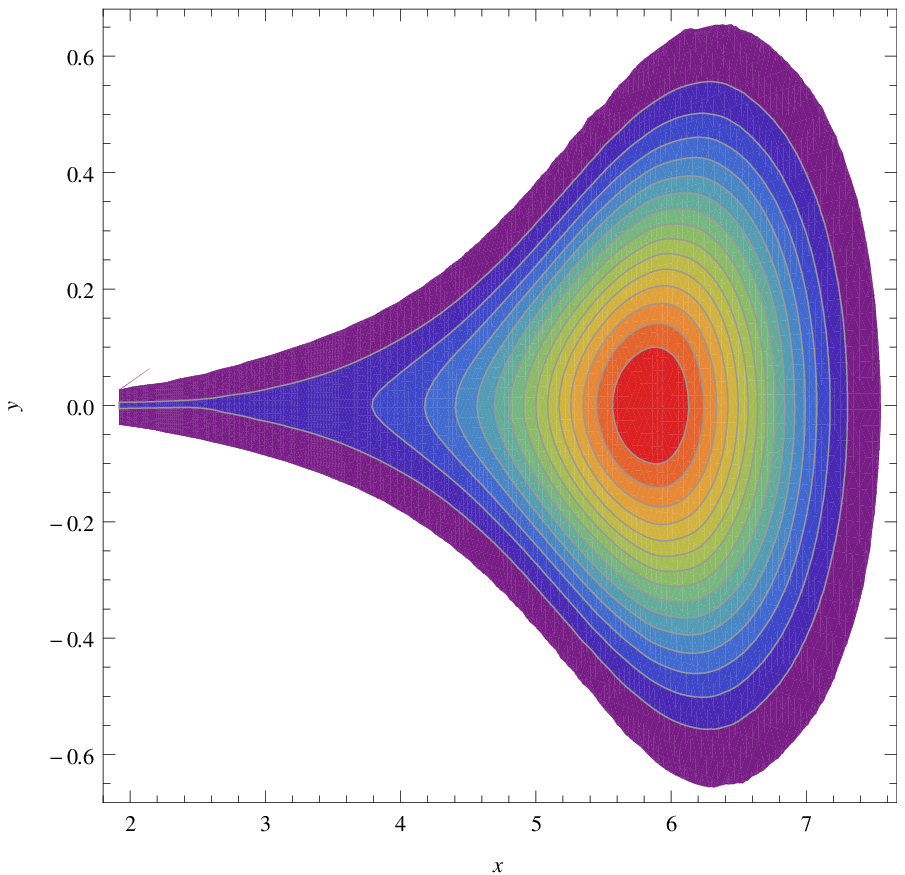}~\hfill~\\[-10pt]~ \hspace*{2cm}$t=0$ \hfill\hspace*{1cm} $t=200$ ~\hfill~\\[20pt]
~\hfill~
\includegraphics[angle=0,width=0.4\textwidth]{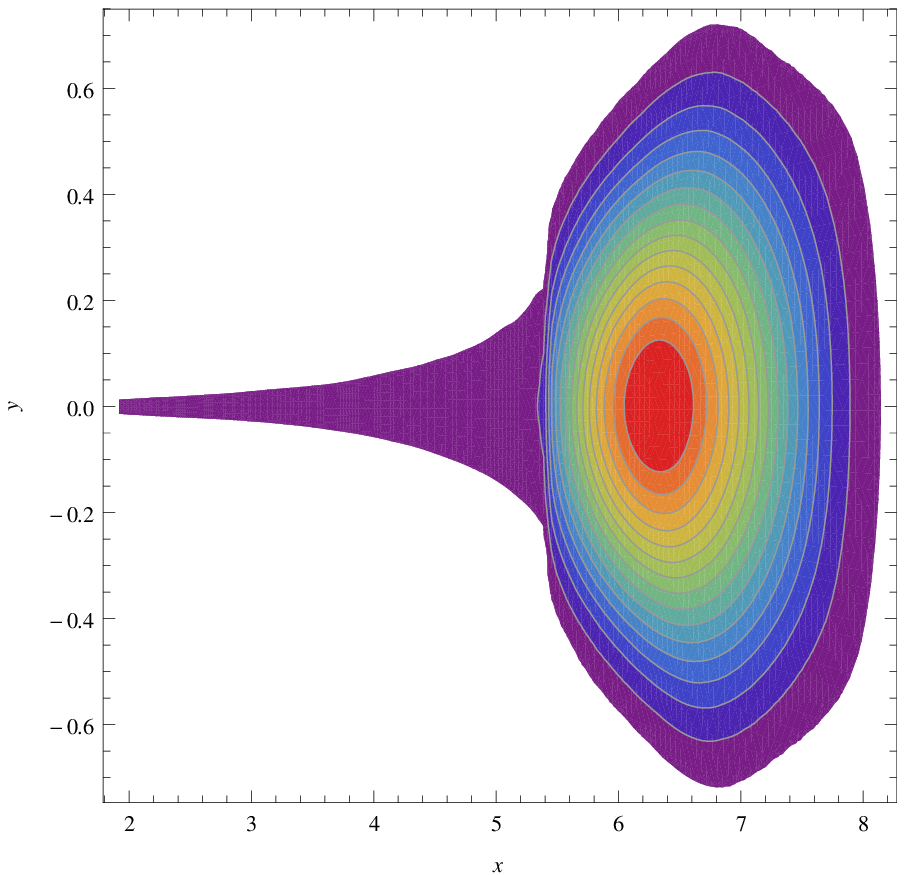}
\hfill
\includegraphics[angle=0,width=0.4\textwidth]{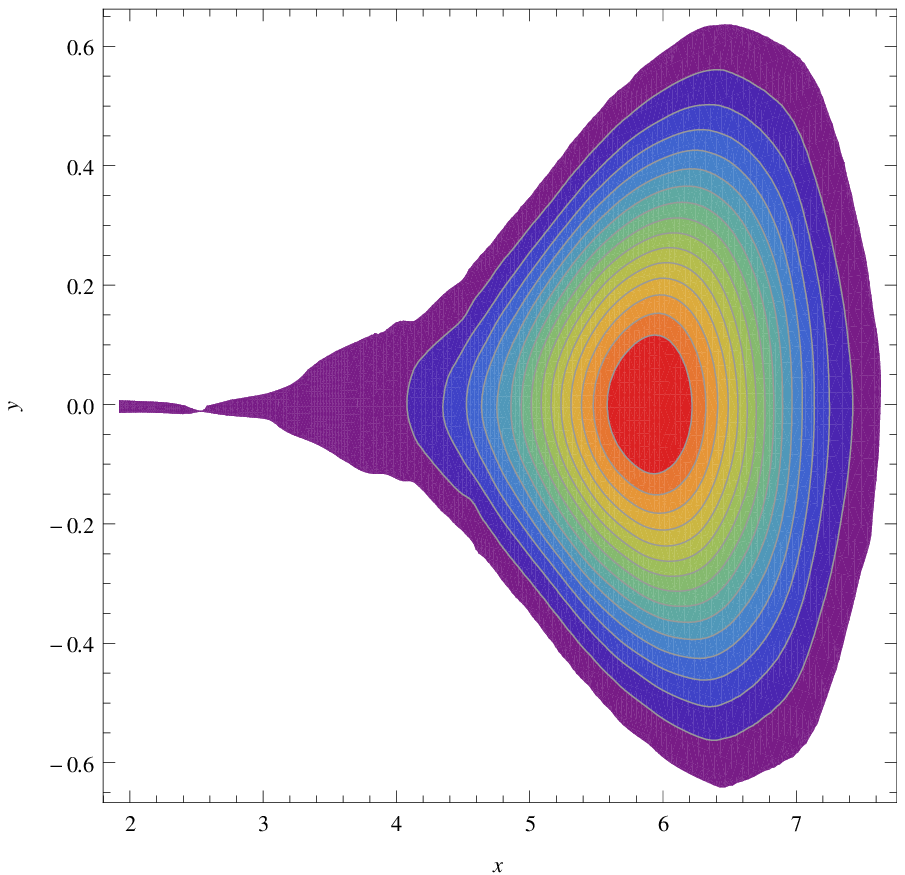}~\hfill~\\[-10pt]~\hspace*{2cm}$t=1000$ \hfill\hspace*{1cm} $t=2000$~\hfill~
\end{center}
 \caption{Time evolution of density levels in four snapshots of the poloidal section 
 ($x,y$) across the accretion torus ($x=0$ is the symmetry axis). 
 This example starts from the critical equilibrium configuration,
 which is perturbed at the initial moment of time. At this moment the Keplerian
 orbital time at the inner cusp, $R=5GM/c^2$, corresponds to $T_{\rm Kep}=70GM/c^3$. 
 Subsequently, accretion across the inner edge
 takes place onto the black hole (the outer horizon is at the left side of the 
 panel). This process proceeds in an oscillatory manner, interchanging the phases of
 fast and diminishing accretion rate. The plasma magnetization parameter is set to 
 $\beta = 25$ (sub-equipartion pressure of the magnetic field), 
 the black-hole spin $a=0.1$ (slow rotation of the black hole) in this example.
 The same colour scale as in Fig.~\ref{Fig1}. Geometrized units are used, where
 the length is scaled with respect to the gravitational radius, $GM/c^2$, and time is scaled by $GM/c^3$. }
 \label{Fig4}
\end{figure*}

An exemplary profile of such a torus is shown in Figure~\ref{Fig1}, where
contours of constant mass density are constructed from the analytical form of the solution 
\eqref{Sol5}. The contours are over-plotted on top of the colour-coded density structure. 
The black hole is located towards the left side of the plot; the
coordinates are defined in such a way that the entire left edge of the plot corresponds to the 
location of the outer horizon at $r=r_+$.
The bottom panel shows the same configuration in Boyer-Lindquist coordinates $(r,\vartheta)$. 
The torus surface extends from the inner rim at $r=4.5$ up to the outer boundary at $r=8.4$
in the equatorial plane, $y=0$. The right edge is at $x= 15$; Cartesian coordinates are derived from 
Boyer-Lindquist coordinates, $x^2+y^2=r^2$ ($x=r\sin\vartheta$, $y=r\cos\vartheta$) and scaled 
with units of $GM/c^2$. The colour 
bar is scaled logarithmically with values of normalised rest-mass density $\rho(x,y)$ relative 
to its maximum value $\rho_{\rm c}$ at the torus centre ($x=R_{\rm c}=6.3$, $y=0$). 
In the bottom panel the horizon is located at $\sqrt{x^2+y^2}=1.95$, so it is hidden behind the left
border of the graph.

We notice that the constant 
density and pressure contours from eq.~(\ref{Sol5}) correspond accurately to the colour scale of the image.
Since the latter is based on a direct output from the numerical code, we can be confident that
the computed structure and the necessary transformations of the coordinates are correct. 
This is a reassuring check before we embark on the time evolution of a perturbed
configuration, where the analytical calculation is not available.

Let us note that the HARM code \citep{gam03} defines radial 
and latitudinal coordinates in a way that helps resolving the plunging region near above the black-hole horizon.
On the other hand, the sharp inner cusp of the critical configuration is revealed more clearly in standard
Boyer-Lindquist coordinates. Figure~\ref{Fig1} compares the torus structure in both types of 
coordinates.

\subsection{Time evolution of perturbed configuration}
We assume that the above-described initial stationary state is pushed out of equilibrium. 
This leads to the capture of a small amount of material by the black hole, which increases the 
black-hole mass, and so the accretion occurs. \citet{abr98} argued that tori with 
radially increasing angular momentum density are more stable. Therefore, we started with
$q>0$ (see Figure~\ref{Fig2}) and concentrated on the influence of the magnetic field on the 
accretion rate. 

The algorithm of the numerical experiment proceeds as follows. At the initial step the mass 
of the black hole was increased by a small amount, typically by about few percent. 
After the time step 
$\delta t$, the elementary mass $\delta M$ and angular momentum $\delta L=l(R_{\rm in})\,\delta M$ 
are accreted across the horizon, $r=r_+\equiv[1+\sqrt{1-a^2}]\,GM/c^2$. 
The mass increase $\delta M$ is computed as a difference of the mass of torus 
$M_{\rm d}= \int_{\cal V}\rho\,d{\cal V}$ at $t$ and $t+\delta t$, where
$d{\cal V}=u^t\sqrt{-g}\,d^3x$ is taken over the spatial volume occupied by the torus.
The corresponding elementary spin increase is $\delta a = l \,\delta M/(M+\delta M)$. 
Therefore, at each step of the simulation we updated the model parameters by the
corresponding low values of mass and angular momentum changes: 
$M \rightarrow M+\delta M$, $a \rightarrow a+\delta a$. The inner cusp moves 
accordingly.

We employed geometrized units, 
setting the speed of light and the gravitational constant equal to unity, $c=G=1$.
This implies the scaling of various quantities with the central black-hole mass $M$. 
However, the mass as well as the spin parameter
evolve gradually (adiabatically) as the accretion of material proceeds from the torus,
$M\equiv M(t)$, $a\equiv a(t)$.
Corresponding quantities in physical units are obtained by the
following conversions:
\begin{equation}
  \frac{M^{\rm phys}}{M^{\rm phys}_{\odot}}=
     \frac{M}{1.477\times 10^{5}{\rm cm}} ,\quad
  a^{\rm phys}=ca, \quad R^{\rm phys}=R.
\end{equation}
Also,
\begin{equation}
  \frac{a}{M}=\frac{a^{\rm phys}}{GM^{\rm phys}/c} \,,\quad
  \frac{R}{M}=\frac{R^{\rm phys}}{GM^{\rm phys}/c^{2}} \,.
\end{equation}
To obtain the frequency in physical units [Hz], one uses the relation
$\kappa^{\rm phys}=c\kappa$.
The geometrized frequencies are scaled by $M^{-1}$. 
Therefore, their numerical values must be multiplied by the factor
\begin{equation}
  \frac{c}{2\pi M}=(3.231\times 10^{4}{\rm Hz})\,
                   \left(\frac{M}{M_{\odot}}\right)^{\!-1}
\end{equation}
to find the frequency in [Hz].

Figure \ref{Fig3} shows the dependence of the torus mass on time for different values $\beta$ 
of the ratio between thermodynamical and magnetic pressure (plasma parameter), 
$\beta \equiv P_{\rm g}/P_{\rm m}$, for a torus with the radially increasing distribution
of angular momentum, 
$l(R)=l_{{\rm K},\,R=R_{\rm in}}[1+\epsilon (R-R_{\rm in})]^q$ with $q>0$, $0<\epsilon \ll1$. 
This means that the reference level of the angular momentum density is set to $l={\rm const}=l_{\rm K}(R_{\rm in})$, 
motivated by the standard theory of thick accretion discs, where the constant value is a limit for 
stability. A radially growing profile then helps to stabilise the configuration.

Unless 
stated otherwise, we set $q=1$, $\epsilon=0.03/l_{\rm}(R_{\rm in})$ for definiteness of
examples in the simulations. At the inner edge of the torus the angular momentum 
equals the Keplerian value, and for higher radii it grows to super-Keplerian rotation,
taking into account the specific shape of the relativistic Keplerian angular momentum
\citep[e.g.,][]{frag2013}. Furthermore, 
following the von Zeipel theorem, in the vertical direction along the constant
$R={\rm const}$ surface within the torus, the angular momentum is defined by its value in the 
equatorial plane. The topology of these surfaces is cylindrical except for relativistic deviations that are
important only at very small radii \citep{chak91}.

From the graph we see that the amount of accreted mass is generally larger for smaller $\beta$. 
The plot also shows that the overall gradually decreasing trend is superposed with fast oscillations. 
After the initial drop of the torus mass (given by the magnitude of the initial perturbation, 
$\delta M\simeq0.01 M$) phases of enhanced accretion change with phases of diminished or zero 
accretion. 

\begin{figure}
 \resizebox{\hsize}{!}{\subfloat{\includegraphics[width=17cm]{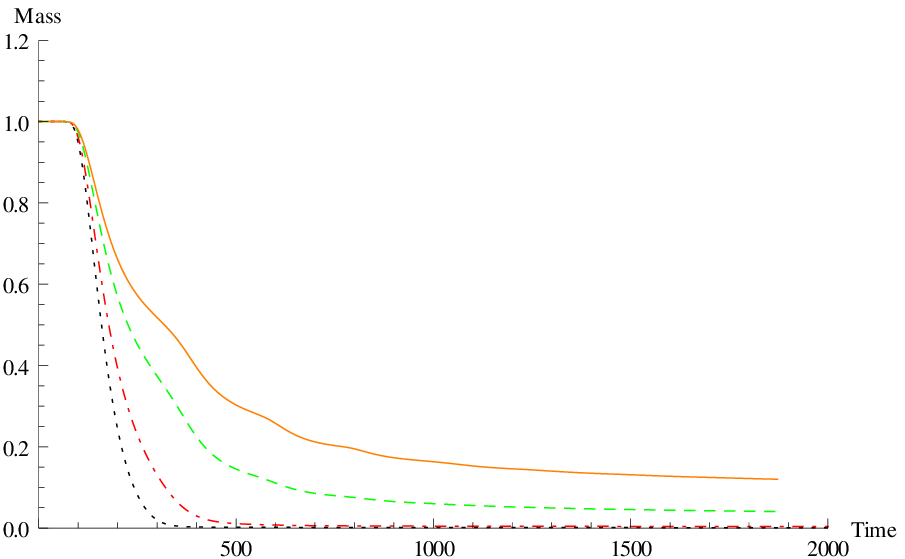}}} 
 \resizebox{\hsize}{!}{\subfloat{\includegraphics[width=17cm]{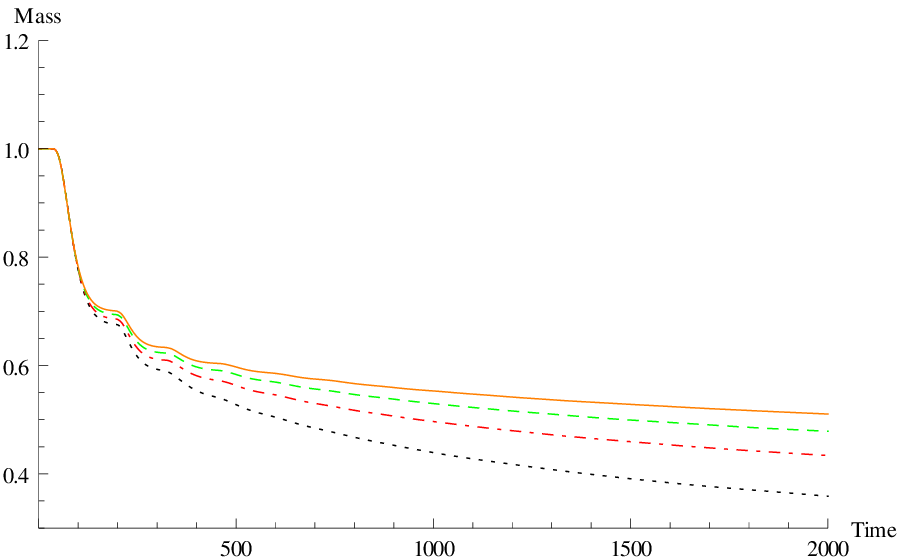}}} 
 \caption{$M_{\rm d}(t)$, mass of the torus, normalised with respect to the starting mass at 
 the initial moment of the simulation, $M_{\rm d}(0)$, is plotted as a function of time for different values 
 of $q=0.6$ (dotted), $0.8$ (dot-dashed), $1.0$ (dashed), $1.1$ (solid). 
 Top panel: $a=0.3$; bottom panel: $a=0.9$. For small $q$ and small $a$ the torus is 
 unstable and its mass becomes quickly accreted onto the black hole, whereas higher values of
 the slope of the angular moment distribution and fast spin of the black hole
 tend to stabilise the system against the initial perturbation (accretion stops at a certain
 moment).}
 \label{Fig5}
\end{figure}

\begin{figure}
  \resizebox{\hsize}{!}{\subfloat{\includegraphics[width=17cm]{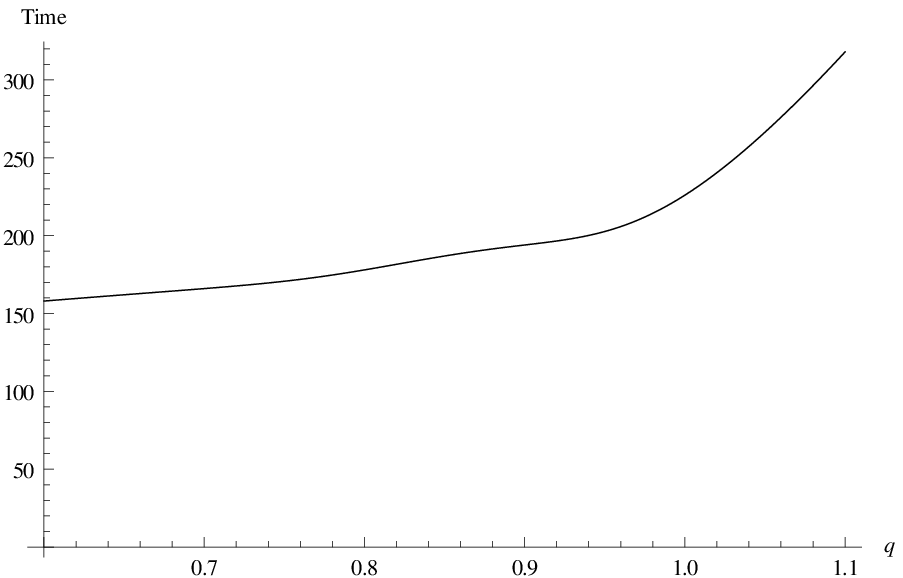}}} \\
  \resizebox{\hsize}{!}{\subfloat{\includegraphics[width=17cm]{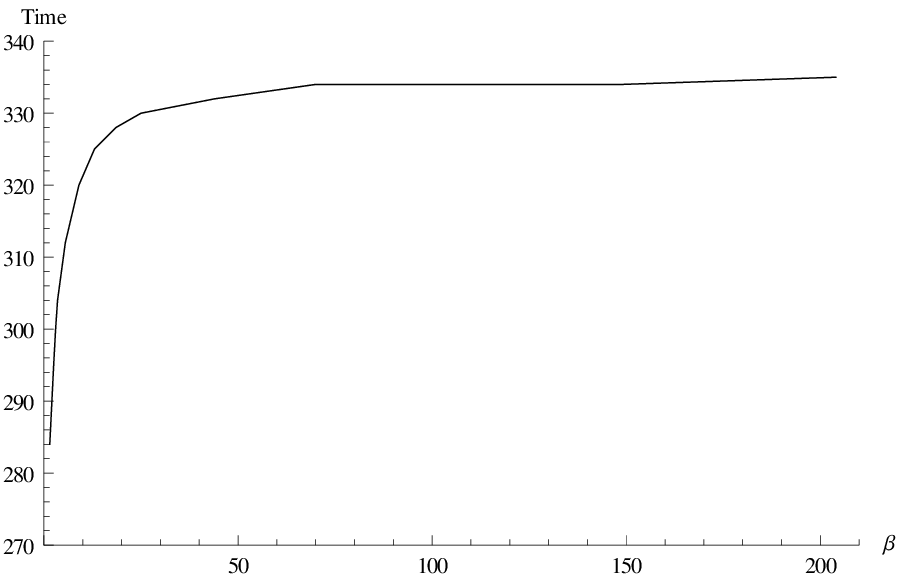}}}
 \caption{Top panel: time to accrete half of the total mass of the torus as a function 
 of $q$  (plasma parameter $\beta=3$). In agreement with the graph of $M_{\rm d}(t)$ in Fig.~\ref{Fig5} we notice 
 a higher rate of accretion (shorter accretion half-time) for lower values of~$q$.
 Bottom panel: the dependence of accretion half-time on the plasma 
 magnetization parameter $\beta$ is shown; the power-law index of the angular moment radial profile 
 is set to a fixed value $q=1.1$.} 
 \label{Fig6}
\end{figure}

\begin{figure}
  \resizebox{\hsize}{!}{\subfloat{\includegraphics[width=17cm]{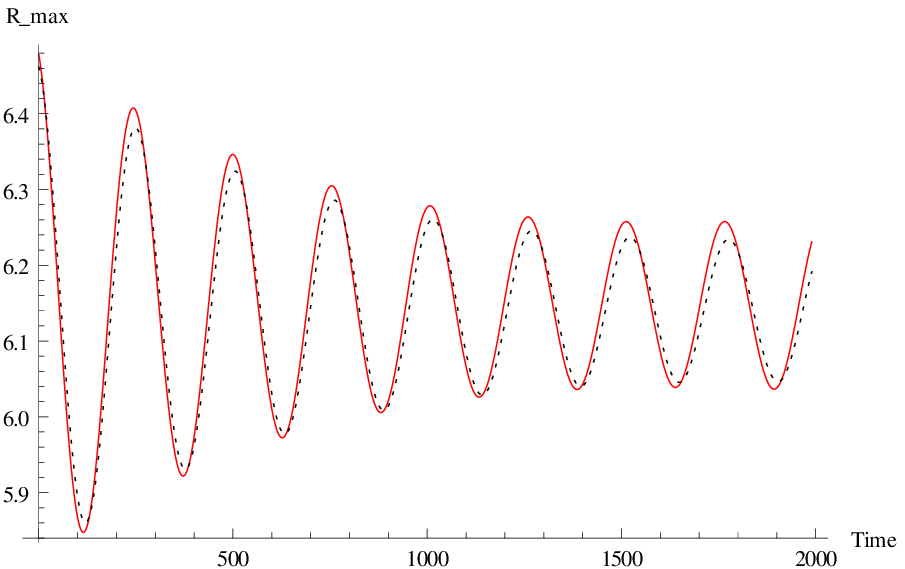}}} \\ 
   \resizebox{\hsize}{!}{\subfloat{\includegraphics[width=17cm]{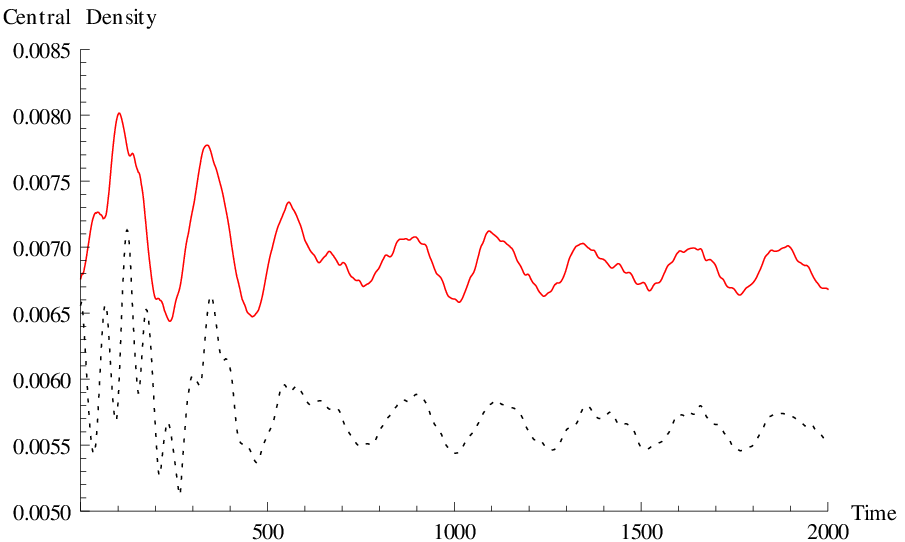}}}
 \caption{Oscillation of the torus centre $R=R_{\rm c}$ (top panel; radius is expressed in geometrized 
 units $GM/c^2$ on the vertical axis), and of the central density $\rho=\rho_{\rm c}$ (bottom panel); 
 density is relative to its peak value at the centre, $\rho_{\rm c}=\rho(R_{\rm c})$. The solid line 
 is for a non-magnetized case ($\beta\gg1$), the dotted line denotes the magnetized configuration 
 ($\beta = 3$).}
 \label{Fig7}
\end{figure}

The oscillatory behaviour can be traced by the position of the torus centre, which we discuss 
below. It resembles the eigenfrequency modes that were proposed as a model for 
quasi-periodic oscillations in some X-ray binaries \citep{rez03,mon12}. A similar accretion history 
is found also for higher spin values, i.e., closer to extreme rotation, although we always assumed 
$|a|<1$ (we did not consider the possibility of a naked singularity, but see \ \citeauthor{stu10},
\citeyear{stu10} for a recent discussion of such a possibility). 

Figure \ref{Fig4} shows levels of mass density at different time moments. Four frames exhibit the 
changing torus structure during the accretion process. (a) The first frame corresponds to the initial equilibrium state; 
no mass overflow takes place and the mass distribution just fills the critical surface. (b) The next frame 
shows the perturbed configuration where the inner edge is pushed slightly outwards, which pushes the
torus out of its initial steady-state. (c) The third frame captures the moment when the accretion drops, 
and finally, (d) in the last frame the mass transfer onto the black hole is completely interrupted, although the
configuration is not exactly stable (accretion is then restored and the cycle continues). We carried out these 
simulations for a different angular momentum dependence on radius to reveal the above-mentioned effect of the 
$l(R)$ profile. 

Figure \ref{Fig5} is complementary to Fig.~\ref{Fig3}. In Figure \ref{Fig5} we compare several cases of different 
angular momentum profiles, as characterised by the slope $q$. One can see that the accretion rate is higher than in the 
previous example. This plot also confirms that for higher $q$ the amount of accreted mass is 
diminished, in agreement with prior studies. In fact, for cases with indices $q=0.6$ and $q=0.8$ almost the whole 
torus is accreted. For $q=1.0$, almost $96\%$ of the initial torus mass is accreted, while for $q=1.1$ it becomes 
$88\%$ of the initial mass. A comparison between the two panels of Fig.~\ref{Fig5} confirms the general trend, which shows 
that tori are more stable for higher spin values, i.e., closer to the extremely co-rotating black hole ($a\rightarrow1$)
for otherwise similar parameters. 
We also followed the oscillations of the point of maximal mass density inside the torus
(the torus centre); it shows a behaviour consistent with the above-described evolution of the accreted mass.

Figure \ref{Fig6} studies the dependence of accretion rate by plotting the half-mass accretion time as a function
of $q$ and $\beta$. Furthermore,
Figure \ref{Fig7} compares the magnetized vs. non-magnetized tori for the same spin ($a=0.3$). In the top 
panel we show the time dependence of the radial coordinate of the point with the highest mass density 
$R=R_{\rm c}$ (hence the highest pressure) of these two tori, and in the bottom panel the dependence of the highest 
mass density is captured as a function of time. In the limit of a non-magnetised slender torus ($R_{\rm c}\gg1$) these oscillations
correspond to the situation that has been treated previously by analytical methods \citep{bla06}. Although 
the amplitude of $R_{\rm c}$ oscillations is quite small
in these examples (because the oscillations were initiated by a weak perturbation and the torus centre is relatively
far from the black hole), the outer
layers of the torus are affected more significantly and can be accreted across the inner edge.

\begin{figure}
  \resizebox{\hsize}{!}{\includegraphics[width=17cm]{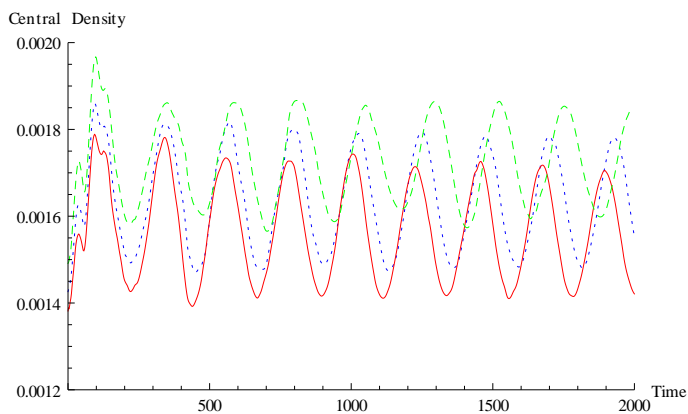}}
 \caption{Central mass density for the fixed angular momentum $a=0.5$ of the black hole
 and varying index of the angular momentum profile of matter in the torus: $q= 0.9$, 
 $1.0$, and $1.1$. Vanishing magnetisation ($\beta\gg1$) in this example.}
 \label{Fig8}
\end{figure}

In these simulations we neglected self-gravitation of the torus \citep{goo88,kar04}. 
However, when the mass and angular momentum are accreted by the black hole, its parameters $M$ and $a$ are 
obviously changed. Hence, even for non-selfgravitating tori we need to update the parameters of Kerr space-time 
metric to achieve a consistent description. In our scheme we changed $M$ and $a$ at each time step, according 
to accreted mass and its angular momentum content. 
Then we can watch 
how this updating influences the accretion process, namely, oscillations of the torus centre, torus mass, and
other characteristics. Figure \ref{Fig8} plots the dependence of the central mass density on time for a 
$l(R) \propto R^q$ profile for the non-magnetized case. In an analogous way, we examined
the role of different initial perturbations $\delta M(t=0)$.

From these illustrations one can deduce that for a sufficiently steep slope $q$ the torus becomes stabilised
with respect to runaway accretion. The steeper $q$, the longer oscillation period. The (weak) 
initial perturbation does not significantly influence the oscillation period, 
only the oscillation amplitude is affected. Qualitatively identical conclusions are obtained for
a slightly different value of the polytropic gas index. As mentioned above, the results presented 
here were computed for $\gamma=4/3$; we also computed the same set of plots for $\gamma=5/3$
with very similar results, while the model is more sensitive to relatively weak variations of $q$.

\begin{figure*}
\includegraphics[angle=0,width=0.48\textwidth]{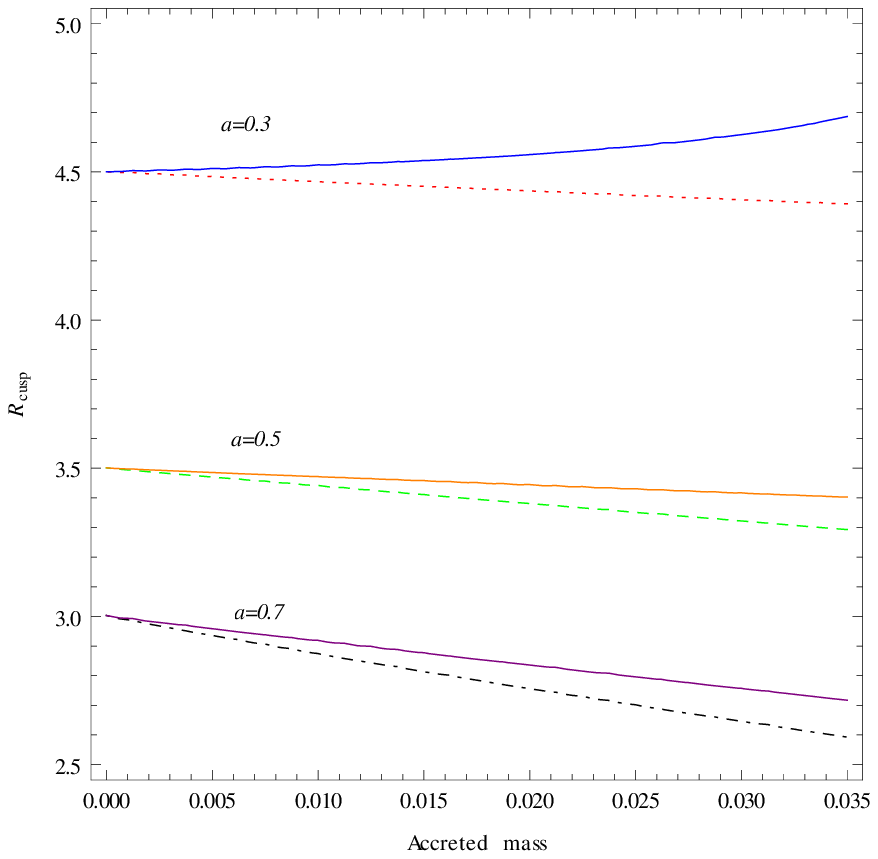}
\hfill
\includegraphics[angle=0,width=0.48\textwidth]{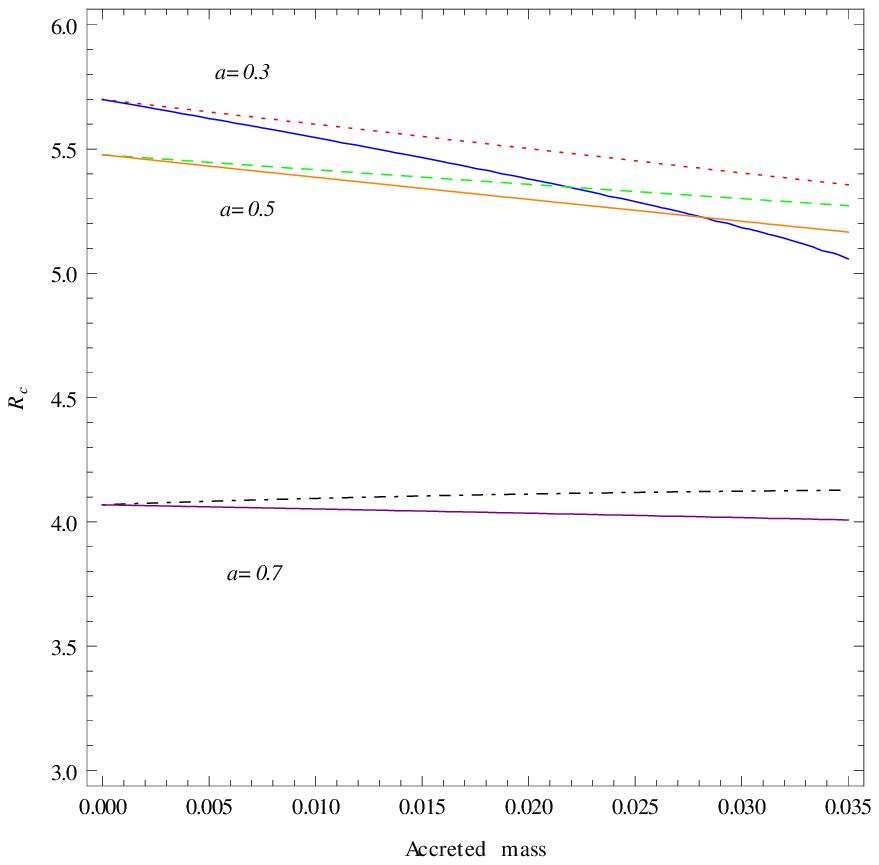}
\caption{Evolution of the characteristic radii during the process of mass transfer from the accretion torus
onto the black hole. Left panel: the mean radius $R=R_{\rm cusp}(t)$ of the cusp (i.e. the inner 
edge of the critical overflowing configuration) as a function of the mass accreted from the torus
onto the black hole, computed by integrating $\Delta M_{\rm d}(t)=\int_{t'=0}^t\delta M(t')$. The Kerr 
metric parameters $M$ and $a$ are evolved during the accretion process (the initial
spin values are given with the curves). 
The solid curve shows the dependence while the material with Keplerian angular 
momentum near the inner edge is accreted; the corresponding broken curve shows this dependence 
for a slightly lower angular momentum than the previous case at the inner edge: the angular-momentum values are 
for $a=0.3\rightarrow l= 3.444\;({\rm resp.\;} 3.15)$, for $a=0.5 \rightarrow l=3.263\;(2.936)$; 
and for $a=0.7\rightarrow l=2.952\;(2.657)$.
The case of growing $R_{\rm cusp}(\Delta M_{\rm d})$
generally corresponds to the receding inner edge, therefore to a shrinking volume of the 
torus, and so gradually increasing rate of mass accretion. Right panel: the dependence
of the torus centre on the accreted mass for the same set of parameters (and
the same notation of line types) as in
the left panel.}
 \label{Fig9}
\end{figure*}

Finally, two additional plots reveal the changing parameters of the torus and the black hole 
in the course of accretion. Figure \ref{Fig9} compares the positions of the torus cusp and the torus 
centre as functions 
of the accreted mass for different initial values of spin $a=0.3,$ $0.5$, and $0.7$.
The evolution of metric parameters $M$ and $a$ has a stabilising effect because the critical 
surface moves inward, so this case corresponds to the situation when the accretion rate and other
characteristics oscillate. It does not lead to the runaway behaviour, which is also reflected in the
gradually decreasing cusp radius. On the other hand, the solid line 
is related to the case when the accreted mass contains less angular momentum, and so the impact of 
increasing the black-hole mass is stronger than the effect of increasing the spin. In this case 
updating the metric parameters results in tori that tend to be more unstable. The mutual relation
between the two radii, i.e. $R_{\rm cusp}$ vs.\ $R_{\rm c}$, provides information about the 
size of the torus as it changes by loosing material onto the black hole and moving in radius,
while the black hole grows.

Naturally, the mean centre of the torus, $R=R_{\rm c}$, moves along with the above-mentioned 
gradual evolution of the inner cusp, $R=R_{\rm cusp}$, by the mass transfer from the torus onto the black hole. 
The torus centre obviously satisfies $R_{\rm c}(t)>R_{\rm cusp}(t)>r_+(t)$ at each moment of the evolution. 
However, the exact mutual relation between these 
radii depends on details of the particular case, namely, the density and the angular momentum distribution 
within the torus. Therefore, both the mean $R_{\rm c}$ and the mean $R_{\rm cusp}$ can either approach 
the centre or recede towards a larger distance, depending on whether the torus shrinks and eventually 
becomes
accreted onto the black hole (which is the case of runaway accretion) or if the partially accreted
structure becomes stabilised against more mass transfer and stays away from the black hole. Naturally,
the centre radius is influenced also by the fact that the black hole itself evolves its mass and spin.
 
\begin{figure}
\resizebox{\hsize}{!}{\includegraphics[width=17cm]{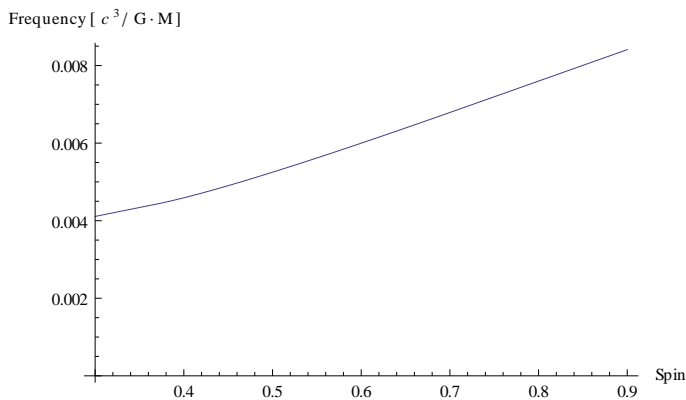}}
\caption{Oscillation frequency of the torus as a function of the black-hole dimensionless 
spin $a$ from the numerical simulation (assuming $0\leq a\leq 1$, $q=1$). Different runs differ from each 
other by the radial position of the centre of the accretion torus, and therefore the oscillation 
frequency also varies. The functional dependence agrees with the radial epicyclic mode.}
\label{Fig10}
\end{figure}

Figure \ref{Fig10} shows the oscillation frequency of the torus, as determined from a sequence of 
our numerical solutions with different position of the torus centre. The frequency
varies gradually in this graph, along with the black-hole dimensionless spin $a$. Again, this change 
can be seen as a result of accretion of the material from the torus onto the black hole, which modifies 
the model parameters including the black-hole spin and the torus centre (as well as the corresponding 
torus mass, $M_{\rm d}$, and other characteristics of the system, as explained above). This dependence allows us to 
unambiguously identify the relevant oscillation mode. We confirm a very precise agreement between the resulting 
curve in Fig.~\ref{Fig10} and the theoretical formula for the radial epicyclic oscillation $\kappa\equiv \kappa(M,a)$
\citep[e.g.][see eq.~(2.105)]{kat08}. The difference between the numerically determined value 
and the analytical formula for the radial epicyclic frequency is less than 1 per cent, and so the two 
dependencies are practically indistinguishable in the plot.

Naturally, to achieve a consistent solution, the parameters need to be evolved during the 
accretion process. Nevertheless, we checked that the above-mentioned point about the metric 
parameters does not influence the oscillation period of a globally stable configuration, i.e., 
until the perturbation itself remains weak. In other words, while the position of 
the torus centre and the magnitude of central density differ at a level of several percent between 
different simulations, the agreement about the oscillation frequency is typically one order of magnitude better.

We recall a useful scheme \citep[see Tab. 1 in][]{fon02} that summarises
a competing role of different agents that influence the stability of geometrically thick accretion tori
near black holes. These are partly real physical effects (such as the angular moment profile of the
accreted material, rotation of the black hole, and self-gravity), and partly reflect the impact of approximations
that are employed to describe the system (such as the Newtonian versus pseudo-Newtonian versus 
general-relativistic models). As mentioned above, radially growing angular momentum and fast
rotation of the black hole tend to stabilise the system against the runaway instability, while self-gravity 
acts instead against stability. Therefore we can also include the magnetic field as another
ingredient into the discussion of stability. However, as demonstrated also above in this paper, 
with the increasing number of different factors and the interplay of mechanisms taken into account, the whole pictures
becomes more complicated than when one had restricted the discussion to the competition of just two or three
degrees of freedom. In the end the outcome of the analysis can depend on detailed values of the 
parameters, e.g. $q$ vs. $a$. Moreover, especially the magnetic field can develop different geometrical structures
on vastly different scales, and so it may be difficult or impossible to characterise the role of magnetic field on
the runaway stability in a simple way.

In our highly simplified scheme we showed that the effect
of the toroidal test magnetic field (embedded in a prescribed manner into the polytropic fluid) just adds to the internal 
pressure of the fluid. Hence it basically enhances the instability in a similar way as any other contribution
that can enhance the pressure above the equilibrium value.

\section{Discussion and conclusions}
\label{discussion}
Within the framework of an axially symmetric magnetized fluid torus model we have extended the 
previous results on the onset of runaway instability of relativistic configurations near a rotating 
black hole. We concentrated on systems with radially increasing angular momentum 
density that are threaded by a purely toroidal magnetic field. We neglected self-gravity of the gaseous 
material (the mass of the torus was set to be at most several percent of the black-hole mass), nevertheless, 
we allowed for a gradual change of the Kerr metric mass and spin parameters by accretion over the inner 
edge. The angular momentum distribution within the torus was also allowed to evolve, starting from
the initial power-law profile. The mass transfer influences the location of the cusp of the 
critical configuration, which can lead to the runaway instability. 

If the profile of the angular momentum increases sufficiently fast with radius
(typically, for $q\gtrsim0.8$), the initial perturbation becomes stabilised by accretion of a small amount 
of material, whereas for small $q$ the instability causes rapid accretion of the torus. The intensity of the 
threaded magnetic field influences the process of stabilisation or destruction of the torus because,
within the framework of the adopted model, the magnetic pressure adds directly to the gas pressure
(plasma parameter $\beta\gtrsim1$).

The process of accretion is not perfectly monotonic, instead, there are changing phases of enhanced 
accretion rate and phases where the mass of torus remains almost constant. The overall gradual 
decrease of the torus mass is superposed with oscillations that can be seen by following the central
density variations on the dynamical time-scale and the position of the centre of the torus. 
The oscillation amplitude is sensitive to the initial perturbation, but the frequency is not, namely, a small 
change of the metric coefficients does not affect the oscillation frequency. 

The 
toroidal magnetic field plays a more important role in the early phases of the accretion process until
the perturbed configuration finds a new equilibrium or disappears because of the runaway instability. 
If the oscillations become stabilised 
with time, no significant differences occur from the corresponding non-magnetized case, 
even when $\beta$ is near unity (equipartition) and the accreted fraction 
of the torus material is significant.

\begin{acknowledgements}
We thank an anonymous referee for helpful suggestions.
We acknowledge support from the student project of the Charles University (GAUK 139810; JH) 
and the collaboration project between the Czech Science Foundation and Deutsche 
Forschungsgemeinschaft (GACR-DFG 13-00070J;  VK).
The Astronomical Institute has been operated under the program RVO:67985815.
\end{acknowledgements}
  
{}
\end{document}